\documentclass{article}
\usepackage{spconf,amsmath,graphicx,hyperref}
\usepackage{booktabs}
\usepackage{amssymb}
\usepackage{pifont}
\usepackage{subfigure}
\usepackage{subcaption}

\title{Gelina: Unified Speech and Gesture Synthesis \\ via Interleaved Token Prediction}
\name{
\begin{tabular}{c}
T\'eo Guichoux$^{1,2}$, Th\'eodor Lemerle$^{2}$, Shivam Mehta$^{4}$, Jonas Beskow$^{4}$,\\
Gustav Eje Henter$^{4}$, Laure Soulier$^{1}$, Catherine Pelachaud$^{1,3}$, Nicolas Obin$^{2}$
\end{tabular}%
\thanks{This work was partially supported by the Wallenberg AI, Autonomous Systems and Software Program (WASP) funded by the Knut and Alice Wallenberg Foundation. This work was performed using HPC resources from GENCI-IDRIS (Grant 2025-[AD011014957R1]).}
}
\address{
$^{1}$ ISIR, Sorbonne Universit\'e, Paris, France\\
$^{2}$ STMS Lab -- IRCAM, Sorbonne Universit\'e, Paris, France\quad{}$^{3}$ CNRS, France\\
$^{4}$ Department of Speech, Music, and Hearing, KTH Royal Institute of Technology, Stockholm, Sweden
}

\begin{document}
\ninept

\maketitle
\begin{abstract}
Human communication is multimodal, with speech and gestures tightly coupled, yet most computational methods for generating speech and gestures synthesize them sequentially, weakening synchrony and prosody alignment. We introduce Gelina, a unified framework that jointly synthesizes speech and co-speech gestures from text using interleaved token sequences in a discrete autoregressive backbone, with modality-specific decoders. Gelina supports multi-speaker and multi-style cloning and enables gesture-only synthesis from speech inputs. Subjective and objective evaluations demonstrate competitive speech quality and improved gesture generation over unimodal baselines.
\end{abstract}
\section{Introduction}
\label{sec:intro}

Human communication is inherently multimodal. Speech and gestures are jointly realized, making speech and gestures coordinated expressions of the same communicative process \cite{kendon1997gesture, mcneill1992hand, Wagner2014GestureSpeech}. Many approaches have been proposed to computationally capture and generate such multimodal dynamics. Important research directions include text-to-speech (TTS) \cite{du2024cosyvoice, valle, lemerle2024linaspeechgatedlinearattention, lemerle24_interspeech}, co-speech gesture generation \cite{Liu_2024_CVPR,lieu_beat_2022, Alexanderson_2023}, human pose synthesis \cite{guo2023momask}, and facial animation \cite{Nyatsanga_2023}. Recent advances in deep learning and the growing availability of multimodal corpora \cite{Liu_2024_CVPR,lieu_beat_2022} have made it possible to envision integrated models that generate several modalities together \cite{tacotronisg,mehta2023diff,mehta2024matchttsg, Mehta_2024_CVPR, fasttalker}. However, most multimodal systems process speech and gestures separately, typically following a cascaded design in which speech is generated first and gestures are added afterward. This strategy is largely driven by the scarcity of large-scale paired corpora and the constraints of working with datasets built in a single modality, either speech or gestures. In such setups, the speech generation process remains unaware of gestural type and timing, leading to weakened synchrony, limited prosodic alignment, and reduced expressiveness \cite{Nyatsanga_2023,mehta2023diff}. Furthermore, this cascaded approach contradicts psycholinguistic evidence showing that speech and gestures are jointly planned in human communication \cite{kendon1997gesture,mcneill1992hand,Wagner2014GestureSpeech}. A unified computational framework not only improves efficiency by giving gesture models direct access to linguistic-prosodic features, but also resonates with psycholinguistic theories of human communication, such as the growth-point hypothesis \cite{Nyatsanga_2023, mcneill1992hand}. Such a framework paves the way for more natural and seamless integration of multimodal behaviors in applications, including Embodied Conversational Agents and social robots.
\\
To address these challenges, we introduce Gelina, a unified framework that simultaneously synthesizes speech and co-speech gestures across multiple voices and gestural styles, using only text as input. Gelina is a discrete, autoregressive model capable of predicting interleaved speech-gesture token sequences. Gestures are subsequently decoded with a gesture flow-matching model \cite{mehta2024matcha}.
\\
Our main contributions are the following:
\begin{itemize}\setlength\itemsep{0.0em}
    \item We introduce the first \textbf{interleaved token autoregressive architecture} for speech-gesture synthesis, aligning modalities within a unified backbone.
    \item We propose a \textbf{training strategy} that leverages large unimodal text-speech datasets to improve generalization under scarce paired data.
    \item We show that Gelina supports \textbf{flexible input modes}: text-only for speech+gesture generation, or text+speech for gesture-only synthesis.
    \item We demonstrate \textbf{bimodal style cloning} (voice and gesture jointly) through sequence continuation, without explicit speaker embeddings.
\end{itemize}
Demonstrations are available at \href{https://TGuichoux.github.io/}{https://TGuichoux.github.io/}.

\section{Background}
\label{sec:background}

\textbf{Co-speech gesture synthesis:}
\label{ssec:bg_gesture}
Gesture generation has recently shifted to data-driven methods \cite{Nyatsanga_2023}. Early approaches used autoregressive sequence modeling to map speech or text to motion sequences \cite{Yoon2020Speech, lieu_beat_2022}, while diffusion-based generators now dominate for their ability to produce detailed, temporally consistent, and natural gestures \cite{Nyatsanga_2023, Alexanderson_2023}. Other works explore discrete motion representations, enabling more controllable synthesis \cite{Liu_2024_CVPR}. These models accept either speech or text as input and typically rely on speaker embeddings for multi-speaker modeling, which limits their generalization ability to speakers unseen during training.
In contrast, Gelina generates both speech and gestures directly from text, and can also clone voice and gestural style through sequence continuation using a speech-gesture prompt, without relying on speaker embeddings.\\
\textbf{Text-to-speech approaches:}
\label{ssec:bg_tts}
Lately, TTS has shifted toward data-driven methods, with notable advances in discrete code modeling \cite{du2024cosyvoice, valle, lemerle2024linaspeechgatedlinearattention}. These systems discretize audio into tokens using large pre-trained codecs \cite{defossez2022highfi, ji2024wavtokenizer} and employ attention-based alignment to map text sequences to speech. Speaker identity is often cloned from a short reference utterance, enabling multi-speaker synthesis. For example, Lina-Speech \cite{lemerle2024linaspeechgatedlinearattention} adopts an autoregressive encoder-decoder with linear attention, where tokenized text is mapped to speech tokens extracted from WavTokenizer \cite{ji2024wavtokenizer} using a position-aware cross-attention mechanism \cite{lemerle24_interspeech}. Such models achieve competitive,  human-like naturalness, but remain unimodal.\\
\textbf{Unified speech and gesture synthesis:}
\label{ssec:bg_uni}
Joint synthesis of speech and co-speech gestures is a relatively recent research direction. Early work such as \cite{maha2010towards} explored rule-based integration of gesture and speech. Tacotron-ISG \cite{tacotronisg} later represented an early neural attempt by extending Tacotron 2 \cite{tacotron2} with a gesture prediction module, though speech and gesture remained only loosely coupled. More recent works have adapted diffusion frameworks to multimodal tasks: in Diff-TTSG \cite{mehta2023diff}, text is mapped via a duration predictor and decoded with two parallel diffusion heads for speech and gesture, while \cite{mehta2024matchttsg} simplified this design with Match-TTSG, by concatenating speech and gesture features and applying a single unified flow-matching head. Both Diff-TTSG and Match-TTSG were restricted to single-speaker settings due to the scarcity of paired multimodal data. Magi \cite{Mehta_2024_CVPR} attempted to address this limitation by augmenting existing corpora with synthetic speech-gesture pairs to enable multi-speaker modeling; however, synthetic data overall quality remains below that of human recordings. In contrast, our model is natively multi-speaker: we pre-train on a large multi-speaker TTS corpus and fine-tune on BEAT2 \cite{Liu_2024_CVPR}, the largest multi-speaker speech-gesture dataset, whereas prior work depended on mono-speaker or synthetic datasets.

\section{GELINA architecture}
\label{sec:method}

\begin{figure*}[t]
    \centering
    \begin{minipage}[t]{0.20\textwidth}
        \centering
        \includegraphics[width=\linewidth]{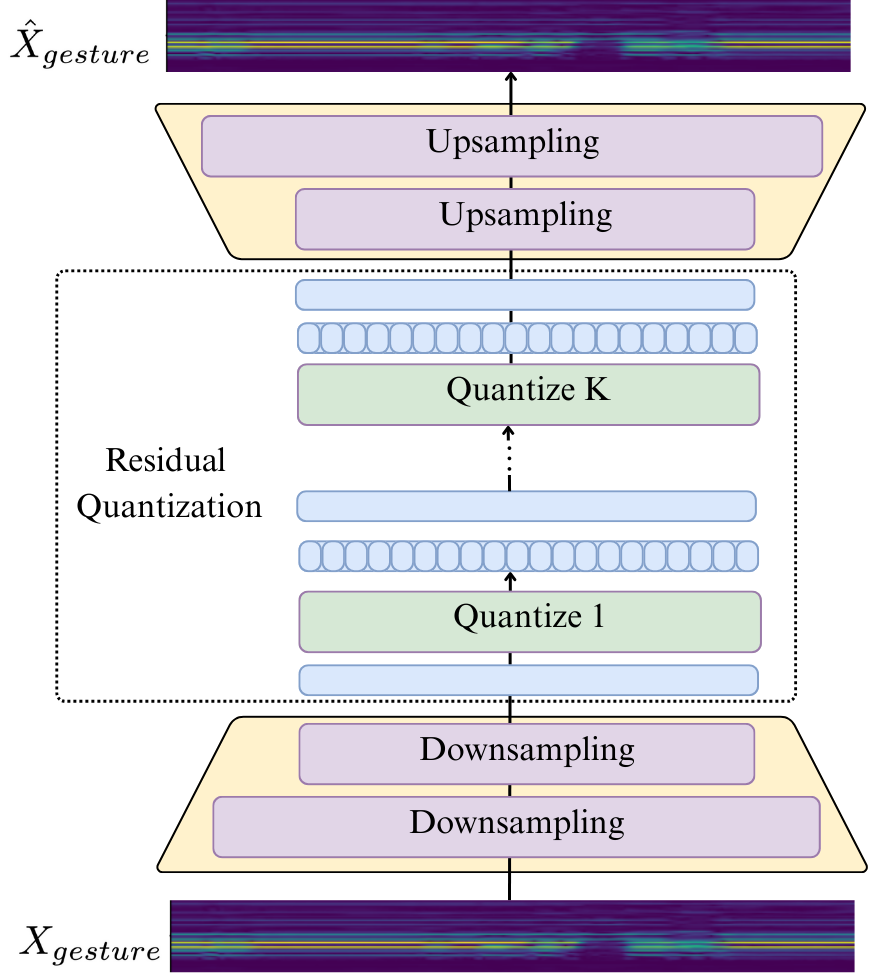}
        \caption*{(a) Gesture tokenizer.}
        \label{fig:rvq}
    \end{minipage}\hfill
    \begin{minipage}[t]{0.55\textwidth}
        \centering
        \includegraphics[width=\linewidth]{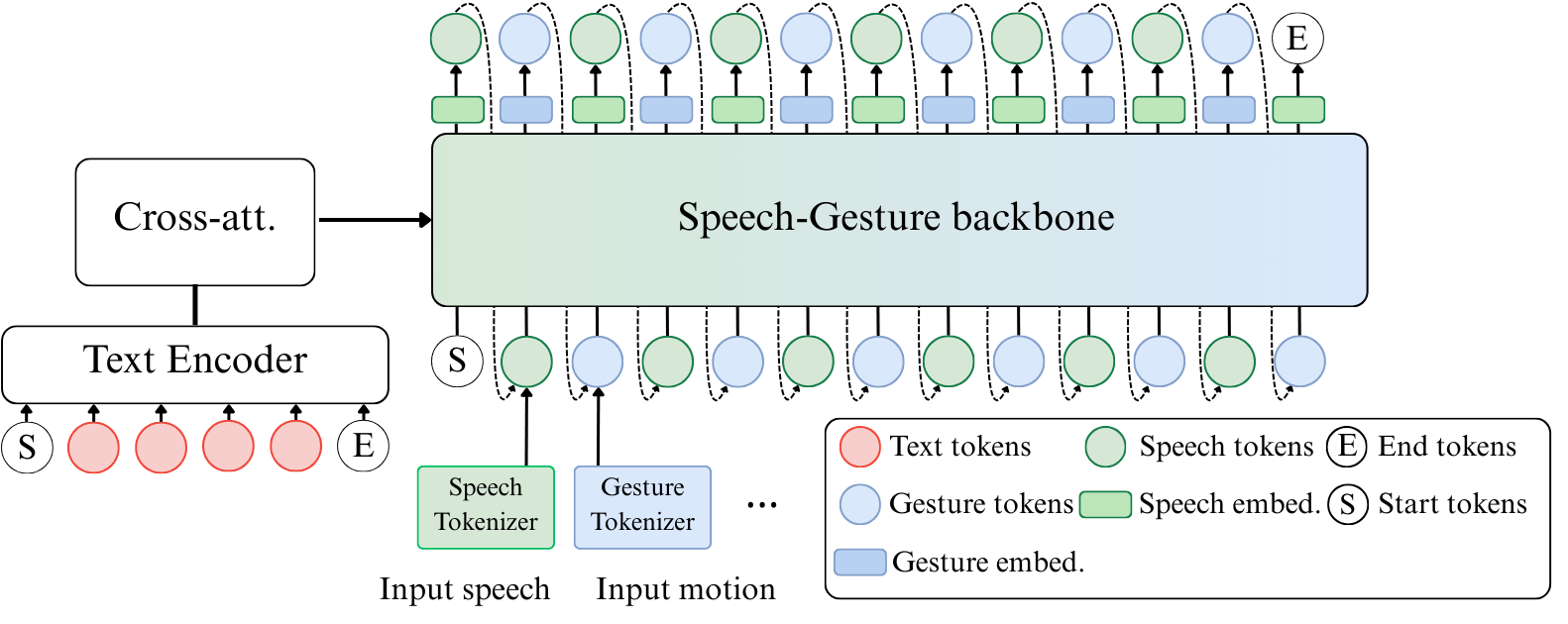}
        \caption*{(b) Speech and gesture autoregressive backbone.}
        \label{fig:backbone}
    \end{minipage}\hfill
    \begin{minipage}[t]{0.25\textwidth}
        \centering
        \includegraphics[width=\linewidth]{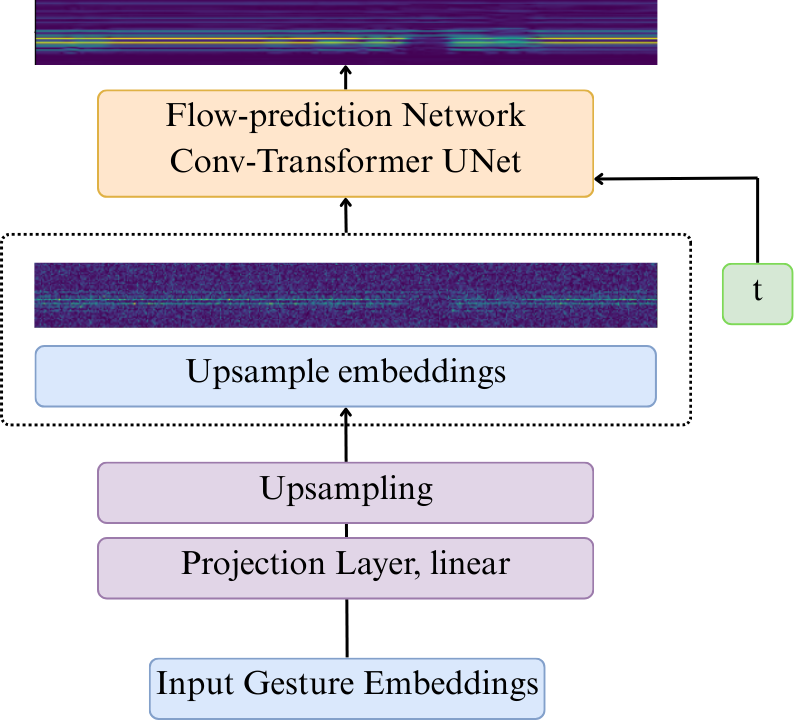}
        \caption*{(c) Gesture conditional flow-matching decoder.}
        \label{fig:cfm}
    \end{minipage}

    \vspace{0.5em}
    \caption{Overview of the proposed architecture: (a) RVQ-VAE gesture tokenizer, (b) autoregressive backbone, and (c) flow-matching decoder.}
    \label{fig:pipeline}
\end{figure*}

Gelina is a bimodal generative model that has three core components, which are depicted for gestures in Figure \ref{fig:pipeline}. The tokenizers independently convert continuous speech and gestures to discrete indices corresponding to latent codes in a vocabulary. The discrete autoregressive transformer temporally aligns text to the sequence of speech and gesture tokens. The decoders convert the outputs of the autoregressive model to the continuous domain using the tokenizer's decoder for speech and flow-matching decoder for gestures.

\subsection{Text, Speech, and Gesture Tokenization}
\label{ssec:tokenization}

 For speech tokenization, we chose to use WavTokenizer \cite{ji2024wavtokenizer}, a discrete audio codec that converts speech waveforms sampled at 24 kHz into discrete tokens at a rate of 75 Hz. Although we observed limitations in WavTokenizer's ability to encode and decode speech from out-of-distribution voices (such as strong accents, noisy speech, or high-pitched voices), we selected it primarily because a single step of quantization suffices to encode the speech signal, unlike popular codecs like Encodec \cite{defossez2022highfi} which require additional residual quantization levels.
 For gesture tokenization, we followed the method proposed by \cite{guo2023momask} and \cite{Liu_2024_CVPR}. We trained a Residual Vector Quantized Variational Autoencoder (RVQ-VAE) to discretize continuous motion sequences into a hierarchical set of discrete tokens. Specifically, the RVQ-VAE first encodes motion into an initial discrete representation and then iteratively encodes the residual error (the difference between the original and reconstructed motion) into additional discrete tokens. This multi-level approach refines the representation at each level, progressively reducing the reconstruction error. We adopted the loss function introduced by \cite{Liu_2024_CVPR}, as it has proven more effective in capturing fine-grained whole-body movements. Our gesture tokenizer is presented in Figure \ref{fig:pipeline}.
For the text tokenization, we used a standard byte-pair encoding algorithm (BPE) \cite{sennrich-etal-2016-neural}.

\subsection{Autoregressive Backbone}
\label{ssec:backbone}

The proposed synchronized speech-gesture generation architecture is based on Lina-Speech \cite{lemerle2024linaspeechgatedlinearattention}, an autoregressive model originally designed for text-to-speech synthesis. Lina-Speech is parameter-efficient compared to other state-of-the-art architectures and integrates linear attention with cross-attention-based text-speech alignment \cite{lemerle24_interspeech}. Gelina introduces an extension of this architecture to multimodal generation of speech and gestures. To do so, a modality-interleaving scheme is proposed: speech and gesture tokens are interleaved by inserting a gesture token every 15 speech tokens. This ratio reflects the encoding rates of WavTokenizer (75 Hz) and Gesture RVQ-VAE (5 Hz). Additionally, separate input embeddings and output projections are maintained for each modality. The complete AR backbone is presented in Figure~\ref{fig:pipeline}.
In practice, the AR backbone is trained with next-token prediction. First, it is pretrained on text-speech pairs from large-scale text-speech datasets, conditioning solely on text transcripts. During pretraining, gesture tokens are replaced by random tokens uniformly sampled from the gesture vocabulary and excluded from the Cross-Entropy loss computation, preserving sequence alignment. Second, the model is fine-tuned on paired text-speech-gesture data to enable synchronized speech-gesture synthesis, still conditioning only on text.
This two-stage training enables the model to establish robust text-speech alignment and, by jointly generating speech and gesture tokens within a single autoregressive stream, ensures synchronized multimodal outputs without relying on external alignment or post hoc gesture prediction. It also allows us to efficiently exploit existing mono- and bi-modal data resources, mitigating the scarcity of large fully paired datasets.

\subsection{Conditional flow-matching}
\label{ssec:cfm}

In preliminary experiments, we observed that decoding gestures directly with the RVQ-VAE decoder yields limited quality, as it is sensitive to noisy gesture token sequences. Moreover, we hypothesize that the AR backbone’s embedding space is semantically richer \cite{betker2023betterspeechsynthesisscaling}, since it integrates multimodal information. To improve decoding, we employ a \textit{conditional flow-matching decoder}, which learns a mapping from noisy gestures to continuous gestures conditioned on the backbone embeddings. The architecture follows Matcha-TTS \cite{mehta2024matcha}, where the flow predictor is a 1D-convolution-transformer UNet. The full gesture flow-decoder is depicted in Figure \ref{fig:pipeline}.

Given a clean gesture sequence $x_0$, a noisy reference sample $x_1$, and a time step $t \sim \mathcal{U}(0,1)$, we define an interpolation path $x_t = (1-t)x_0 + t x_1$ with target vector field $u_t = x_1 - x_0$. The conditional flow-matching objective is  

\[
\mathcal{L}_{\text{FM}}
=
\mathbb{E}_{x_0, x_1, t}
\left[
\big\|
v_\theta(x_t, t, c) - u_t
\big\|_2^2
\right],
\]

\noindent where $c$ are the AR backbone embeddings. To better capture motion dynamics, we extend this objective with a \textit{velocity consistency term}: 

\[
\mathcal{L}_{\text{vel}}
=
\mathbb{E}\!\left[
\|\Delta g_\theta(x_t, t, c) - \Delta x_0\|_2^2
\right],
\]

\noindent and a \textit{geodesic loss} over joint rotations,  

\[
\mathcal{L}_{\text{geo}}
=
\mathbb{E}\!\left[
\frac{1}{KT}\sum_{k=1}^{T}\sum_{j=1}^{K}
d_{\mathrm{SO}(3)}\!\big(\hat{R}_{k,j}, R_{k,j}\big)^2
\right],
\]

with $d_{\mathrm{SO}(3)}$ the geodesic distance on $\mathrm{SO}(3)$. The final training objective is a weighted combination: 

\[
\mathcal{L}
=
\mathcal{L}_{\text{FM}}
+ \lambda_{\text{vel}} \mathcal{L}_{\text{vel}}
+ \lambda_{\text{geo}} \mathcal{L}_{\text{geo}}.
\]

\section{Experiments}
\label{sec:experiments}

\subsection{Experimental setting}
\label{ssec:experimental_setting}
We pre-trained Gelina on GigaSpeech \cite{Chen_2021}, LibriTTS \cite{zen19_interspeech}, and MLS-10k \cite{Pratap_2020}, totaling 18.19k h.
We then fine-tuned our model on the BEAT2 dataset \cite{Liu_2024_CVPR}, which contains aligned speech, gesture, and text sequences. Because of inconsistencies in the provided transcriptions, we re-transcribed the audio using Whisper-large-v3 (temp=0) \cite{whisper}.
Gesture sequences are represented as SMPL-X motion sequences \cite{smplx}, with 55 joints represented as 3D axis-angles rotations. Following \cite{Liu_2024_CVPR}, these pose sequences are converted to Rot6D \cite{rot6d}, with additional translation and foot contacts information. In our experiments, we removed the joints corresponding to fingers to facilitate training, resulting in motion sequences in $\mathbb{R}^{T\times(25_{\text{joints}}\times6+4_{\text{feet}}+3_{\text{trans}})}$. \\
\textbf{Data.} It should be noted that the text-speech datasets used for pre-training are primarily composed of audiobook or read speech \cite{zen19_interspeech}, whereas BEAT2 contains more spontaneous speech and includes several non-native English speakers with noticeable accents.\\
\textbf{Gesture tokenizer.} Our RVQ-VAE has 6 residual layers, 512-entry codebooks, and a latent dimension of 512. Input gesture sequences are at 20 fps and are downsampled by temporal convolutions to 5 Hz embeddings (see \cite{guo2023momask} for architecture). We trained the RVQ-VAE for 90k steps on 1×A6000.\\
\textbf{AR backbone.} The backbone is a 168M-parameter encoder-decoder Transformer with linear attention: a 6-layer text encoder, a 12-layer causal decoder, and a 1024-dimensional latent space. The speech and gesture vocabularies contain 4096 and 512 tokens, respectively; other hyperparameters follow \cite{lemerle2024linaspeechgatedlinearattention}. We observed that discarding motion residuals significantly stabilizes training, so we keep a single residual-quantization level for gestures and rely on the flow-matching decoder to recover the dropped detail. Pre-training runs for 100k steps on 4×H100 with 60k tokens per batch and a learning rate of $2\times10^{-4}$; fine-tuning runs for 5k steps on 1×H100 with 15k tokens per batch and a learning rate of $5\times10^{-5}$.\\
\textbf{Gesture decoder.} We use an 11.5M-parameter U-Net with Matcha-TTS hyperparameters \cite{mehta2024matcha}, plus a projection from gesture embeddings ($d{=}1024$) to continuous motion ($d{=}157$). The flow-matching model is trained for 300k steps on 3×H100 with $\lambda_{\text{vel}}{=}0.05$ and $\lambda_{\text{geo}}{=}0.8$. At inference, we set the number of sampling steps to 100, which we empirically found to balance speed and output quality.

\subsection{Evaluation and results}
We evaluate Gelina on four aspects: (i) speech quality, (ii) gesture quality, (iii) speech-gesture synchrony, and (iv) voice similarity, with unimodal baselines as references. We also assess cloning mode, which transfers voice and gestural style.\\
We selected three baselines for gesture evaluation: \textit{CAMN }\cite{lieu_beat_2022}, \textit{EMAGE} \cite{Liu_2024_CVPR}, and \textit{RAG-Gesture (RAG)} \cite{mughal2024raggesture}. CAMN and EMAGE were retrained with their original implementations to support multi-speaker voice input using a speaker embedding. For the evaluation of the speech modality alone, we considered two baselines: \textit{Lina-Speech} \cite{lemerle2024linaspeechgatedlinearattention} and \textit{CosyVoice-2} \cite{du2024cosyvoice}. CosyVoice-2 uses a 0.5B-parameter text-speech LM trained on over 168k h of multilingual data. We evaluated CosyVoice-2 with the released checkpoints in a zero-shot voice cloning setting. Our Lina-Speech baseline matches Gelina in size ($\sim$168M) and data (18.19k~h) but omits gesture tokens. \\
This comparison setup is somewhat unfavorable to Gelina, since unimodal baselines are specialized for a single modality. Achieving competitive results against specialized unimodal baselines underscores the strength of a unified approach.\\
While several speech-gesture generation models exist \cite{mehta2024matchttsg, Mehta_2024_CVPR, mehta2023diff, tacotronisg}, they rely on upper-body pose representations and are trained exclusively on the Trinity Speech-Gesture dataset \cite{trinity}. These constraints make them poorly suited for our setting, which targets full-body synthesis in SMPL-X. Adapting such models would require substantial redesign and re-training, and their performance would remain non-comparable due to their specialization for Trinity. We therefore focus our evaluation on BEAT2, the largest SMPL-X dataset with multiple speakers, which provides a more appropriate and scalable benchmark.

\noindent\textbf{Objective evaluation}. We evaluate all systems' gesture generation using three standard metrics: Fréchet Gesture Distance \cite{Yoon2020Speech} restricted to body gestures (FGD-B), which measures the distance between the distribution of generated gestures and that of human-recorded gestures; Beat Consistency (BC) \cite{li2021learn}, which assesses temporal alignment between gesture beats and audio beats; and L1-Diversity (Div.) \cite{Liu_2024_CVPR, li2021audio2gestures}, which quantifies the variability of generated gesture sequences. As our evaluation does not account for finger motion, all finger joints were zeroed before computing Div. and FGD. \footnote{To match prior work that reports full-body metrics, we also compute BEAT2 full-body FGD and Div (hands included) for the gesture baselines. Relative to human (FGD = 0.0; Div. = 14.31): CAMN 0.501 / 9.90, RAG 0.538 / 16.65, EMAGE 0.841 / 11.15.}

Speech synthesis quality is assessed with Word Error Rate (WER), reflecting intelligibility, Natural MOS (NMOS) \cite{Mittag_2021}, a predictor of naturalness, and Speaker Similarity (SS), which measures the similarity between generated speech and reference speakers. WER is computed with Whisper-large-V3 \cite{whisper}, and SS follows \cite{du2024cosyvoice}, using cosine similarity of WavLM-large \cite{wavlm} embeddings. 
We evaluated Gelina under three configurations: \textit{Gelina}, the base model; Gelina Cloning (\textit{Gelina clon.}), where sequence continuation is conditioned on an additional speech-gesture input to clone both the speaker’s voice and gestural style; and \textit{Gelina S2G}, where ground-truth speech is provided as input and only the generated gesture sequence is assessed. We included \textit{Gelina - Flow}, an ablation of \textit{Gelina Clon.} without the flow-matching decoding, as well as \textit{Tokenizers}, which reconstruct speech and gesture directly from the codecs without autoregressive modeling.
Table \ref{tab:results} reports objective evaluation results on the BEAT2 dataset.

Among gesture metrics, Gelina Cloning achieves the lowest FGD-B, indicating closest match to human distributions, and competitive BC and diversity scores compared to baselines. Gesture-only models such as RAG obtain closer beat consistency and higher diversity, but at the cost of higher distributional distance. For speech, Gelina Cloning reduces WER compared to Lina-Speech and increases NMOS.  Gelina Cloning further approaches CosyVoice-2 in terms of similarity, while maintaining multimodal capability. These results demonstrate that Gelina achieves a favorable balance between gesture fidelity and speech quality, outperforming unimodal gesture baselines and remaining competitive with strong speech-only systems.
\noindent\textit{Runtime.} On a GPU A5000, RTFs (mean over 30 clips) are 1.47 for Gelina, 1.26 for Lina-Speech, and 0.50 for CosyVoice-2. While CosyVoice-2 and Lina-Speech are faster, they are speech-only; Gelina remains near real-time despite synthesizing both speech and gestures.

\begin{figure}
    \centering
    \includegraphics[width=1.\linewidth]{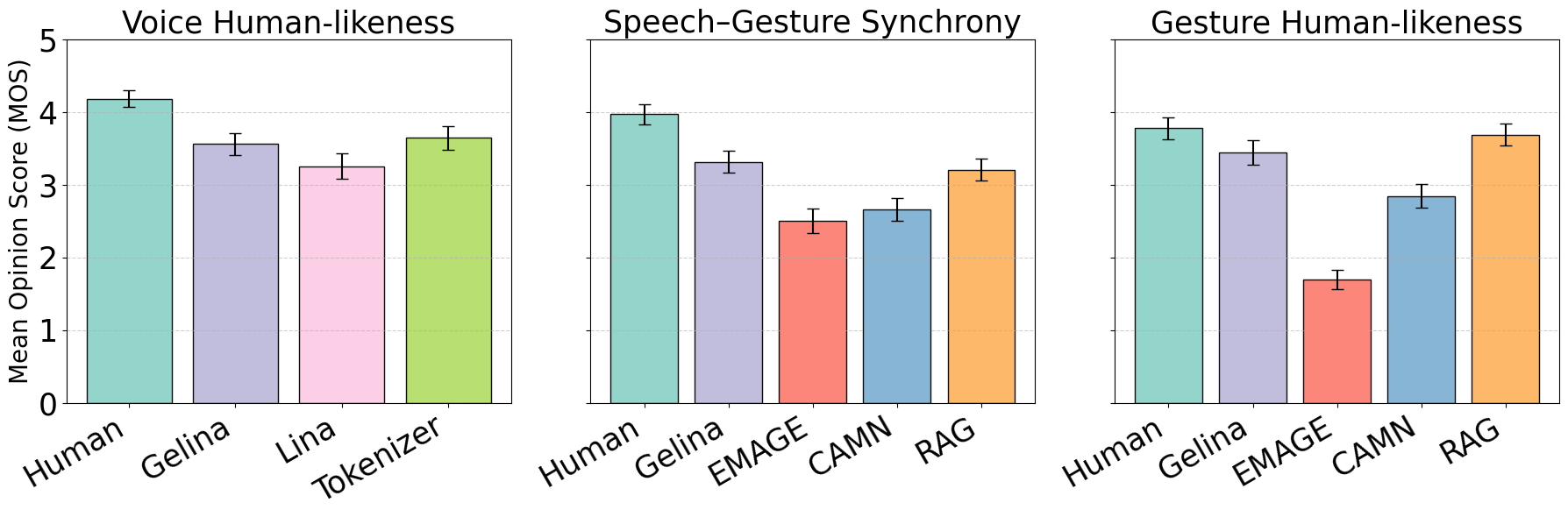}
    \caption{Mean Opinion Scores with 95\% confidence intervals from user evaluation across three aspects: Voice human-likeness, speech-gesture synchrony, and Gesture human-likeness.}
    \label{fig:mos}
\end{figure}

\noindent\textbf{User study}. While objective metrics provide useful indicators of model performance, user studies remain the gold standard for assessing speech and gesture generation. We conducted a large-scale user study with 96 participants, divided into three independent tasks: (i) rating the human-likeness of generated speech (audio only), (ii) rating the human-likeness of generated gestures (animation only), and (iii) rating the synchrony between speech and gestures (audio-visual). Each participant rated 30 stimuli with durations between 8 and 15 seconds and completed 2 attention checks. Participants were recruited through the Prolific platform and received an average compensation of £2.10 per test. Ratings were collected on a 5-point Likert scale. We report Mean Opinion Scores (MOS) with 95\% confidence intervals in Figure~\ref{fig:mos}. For this study, Gelina is evaluated in the cloning setting.\\
Pairwise Student’s t-tests (with one-way repeated measure ANOVA ($p < 0.05$) and Holm correction) show that Gelina significantly outperforms Lina-Speech in voice human-likeness, with scores not significantly different from the speech tokenizer, suggesting the tokenizer as the main bottleneck. The gain over Lina-Speech may come from gesture conditioning, which adds context for emphasis and cloning, or simply from more extensive hyperparameter tuning in Gelina. For synchrony and gesture human-likeness, Gelina and RAG both significantly outperform EMAGE and CAMN; synchrony scores between Gelina and RAG are not significantly different, though RAG rates marginally higher for gesture human-likeness. Gelina achieves this level of performance while also synthesizing speech, demonstrating that competitive gesture quality can be obtained within a unified speech-gesture generation model.

\label{ssec:results}

\begin{table}[t]
\centering
\caption{Evaluation results on BEAT2 (all speakers). 
FGD-B: Fréchet Gesture Distance (Body), BC: Beat Consistency, Div: L1-Diversity, 
WER: Word Error Rate (\%), NMOS: Neural MOS, SS: Speaker Similarity (×100).
\textbf{Bold} = best, \underline{underlined} = 2nd best. WER and NMOS are reported with 95\% confidence intervals}
\label{tab:results}
\footnotesize
\setlength{\tabcolsep}{3pt}
\begin{tabular}{lcccccc}
\toprule
\textbf{Model} & \textbf{FGD-B} $\downarrow$ & \textbf{BC} $\sim$ & \textbf{Div.} $\sim$ & \textbf{WER} $\downarrow$ & \textbf{NMOS} $\uparrow$ & \textbf{SS} \\
\midrule
Human           & 0.0 & 0.684 & 4.14 & 6.5 ±.54 & 3.72 ±.04 & 69.1\\
Tokenizers      & 0.0118 & 0.667 & 3.91 & 11.03 ±.7 & 3.19 ±.04 & 66.8\\
\midrule
CAMN            & \underline{0.1097} & 0.551 & 2.96 & - & - & -\\
EMAGE           & 0.1679 & 0.766 & \underline{3.92} & - & - & -\\
RAG     & 0.1781 & \textbf{0.700} & 5.13 & - & - & -\\
\midrule
Gelina          & 0.2310 & 0.744 & 3.20 & 11.3 ±1.0 & 2.96 ±.04 & - \\
Gelina Clon.  & \textbf{0.0839} & \underline{0.738} & 3.15 & \underline{9.2 ±.84} & \underline{3.21 ±.04} & \underline{61.3} \\
Gelina S2G & 0.1950 & 0.768 & \textbf{4.03} & - & - & -\\
\midrule
Gelina - Flow            & 0.6107 & 0.824 & 4.28  & 9.2 ±.84 & 3.21 ±.04 & 61.3\\
\midrule
Lina-Speech  & - & - & - & 10.9 ±.9 & 2.98 ±.05 & 60.1\\
CosyVoice-2 & - & - & - & \textbf{3.5 ±.5} & \textbf{3.70 ±.04} & \textbf{63.9}\\
\bottomrule
\end{tabular}
\end{table}

\section{Conclusion and future work}
\label{sec:conclusion}
We have presented Gelina, a model for joint speech-gesture generation. We evaluated it through both objective metrics and a user study. Gelina significantly outperforms two gesture baselines, EMAGE and CAMN, and reaches performance comparable to the strongest system, RAG-Gesture, while also delivering competitive speech quality relative to speech-only systems such as CosyVoice-2 and Lina-Speech. Our findings demonstrate that joint speech-gesture generation can remain competitive, and even outperform unimodal baselines, despite the added complexity of synthesizing two modalities. Current limitations are that Gelina models only body gestures, and its speech quality is constrained by the tokenizer. Future work will address these limitations by improving the tokenizer, extending gesture coverage to fingers and facial expressions, and supporting longer sequence generation.

\section{Compliance with Ethical Standards}
This is a numerical simulation study for which no ethical approval was required. 
All datasets were used under license, and the user study was conducted with informed consent and fair participant compensation. Cloning experiments were restricted to consented voices, and demos are released with safeguards to mitigate potential misuse.

\bibliographystyle{IEEEtran}
\bibliography{strings,refs_short}

\end{document}